# Flux Variation of Cosmic Muons

N. Ramesh, M. Hawron, C. Martin, A. Bachri

*Southern Arkansas University, Magnolia, AR 71754*

Correspondence: agbachri@saumag.edu

**Abstract**

In the current paper, we analyzed the variation of cosmic radiation flux with elevation, time of the year and ambient temperature with the help of a portable cosmic muon detector, the construction of which was completed by a team from Southern Arkansas University (SAU) at Lawrence Berkeley National Laboratory (LBNL). Cosmic muons and gamma rays traverse two synchronized scintillators connected to two photomultiplier tubes (PMT) via light guides, and generate electronic pulses which we counted using a Data Acquisition Board (DAQ). Because muons are the product of collisions between high-energy cosmic rays and atmospheric nuclei, and therefore shower onto earth, the scintillators were arranged horizontally for detection. The elevation measurements were recorded at different locations, starting from 60 feet below sea-level at the Underground Radiation Counting Laboratory at Johnson Space Center, TX, to 4200 feet at Mt. Hamilton, CA. Intermediate locations included sea-level Galveston Bay, TX, and Mt. Magazine, AR (2800 feet). The data points showed a noticeable increase in flux as elevation increases, independent of latitude. Measurements investigating the dependence of cosmic rays on temperature and time of the year took place locally in Magnolia, AR. We found that cosmic muon flux is uniform, appears to be independent of conditions on earth, and is anti-correlative with temperature. We are convinced that the sun has minimal to zero effect on cosmic-ray flux; it cannot be a major contributing source of this background radiation. The source of cosmic radiation remains one of the biggest unanswered questions in physics today.

**Introduction**

Cosmic rays (CR) are energetic particles that constantly rain through the Earth's atmosphere, a fraction of which penetrates the Earth's surface at high relativistic velocities. CR are the source of a uniform background ionizing radiation, they shower on the Earth's atmosphere at a rate of about 1000 collisions per square meter per second. The source of these particles varies from the sun to yet unidentified astronomical events in the extremities of the universe, such as supernovae and black holes. CR have not only revolutionized astronomy and particle physics; they may have also played a vital role in human life. Since it is evident that ionizing radiation is indeed mutagenic, it is of significant interest to scientists that cosmic radiation had some responsibility in the evolution of life on our planet; that is cells may have developed an adaptive reaction to these showering particles (Rachen et al. 1993). In addition, gamma rays (GR), the uncharged component of CR, continue to have an integral role in evolution in our planet by inducing double-strand DNA breaks in human cells (Francis et al. 2006). Although the growth in the research today concerning these particles is continuing, the origin of CR and the mechanism by which they are accelerated in the universe is still somewhat of a mystery.

Most of the CR energy arrives to the Earth's surface in the form of kinetic energy of muons. The muon ($\mu^-$ and $\mu^+$) is a particle belonging to the lepton family, and as such, it has the same charge as that of an electron. It is the second-heaviest lepton with mass 105.6 MeV, which makes it 207 times the mass of an electron. Muons are secondary products of interactions between highly energetic CR and the nuclei of atmospheric particles. They are the result of decay of pions ($\pi^-$ and $\pi^+$). Because of their ultra-relativistic nature, the muons created in the atmosphere can permeate the Earth's surface for hundreds of meters. The flux of these particles can be sufficiently distinguished with a scintillator detector system.

A scintillator is a material that becomes luminescent when ionizing radiation is present. In other words, it releases photons due to interaction with a penetrating radiation. These photons are steered, via a plastic light guide, toward the photocathode surface of a photomultiplier, causing the release of electrons by the photoelectric effect. Each electron is multiplied to many more via an amplification process within the PMT achieved by a series of dynodes kept at high voltages. Therefore, a PMT converts light signals into electrical pulses. A circuit board processes the signals from the PMTs, translating them into counts (or data), which are subsequently collected and read through a USB interface. The detector was synchronized with a data acquisition board (DAQ) to enhance the collection of data at various control voltages. The detector is proficient and sensitive enough to collect both GR and CR muon flux information. To ensure the overall consistency and functionality of the machine, we previously (Bachri et al. 2011) investigated the radiation of gamma rays from an active cobalt-60 sealed source, and showed that radiation falls as $1/r^2$, where r is distance from radioactive source, consistent with the inverse square law.

The current study, however, focuses on cosmic muon flux variation with elevation, in addition to temperature and time of the year. Data acquisition spanned over a period of one year, and was taken in different locations. Locations reported here include (with elevations): Galveston, TX (10ft), Magnolia, AR (338ft), Mt. Magazine, AR (2753ft), Mt. Hamilton, CA (4200ft) and inside the Radiation Counting Laboratory (RCL) at NASA Lyndon B. Johnson Space Center (elevation 60ft below sea-level). Measurement clearly showed a dramatic increase in muon flux (counts/min) as the elevation was increased, that is, the closer one gets to the atmosphere, the higher is the count. The data taken at RCL were especially interesting. RCL at Johnson Space Center is a lab 60 feet underground that was especially and originally constructed for low-level counting of returned lunar samples (Keith 1979). The walls of the room are made of three-eighths-inch mild steel plate that was selected for its low radioactivity. The plates are welded together to form an air-tight wall. Three feet of crushed, washed gravel (with very low radioactive content) perfectly encloses the room, shielding it from the natural radioactivity in the outside concrete and soil. Even so, muons were detected inside RCL, clear evidence that they penetrate large amounts of material due to their large energy before finally decaying, or losing their energy due to their ionizing ability. Their lifetime of 2.2 μs, which is relatively large, also allows them to decay relatively slowly and reach the Earth's surface, even penetrating materials they encounter. The trip of muons through the Earth's atmosphere can only be sufficiently understood within the framework of Special Relativity. Indeed, muons are often described as the penetrating component of CR (Gaisser 1990).

Based on time and temperature measurements at different times of the year, it is clear to us that most of the muon-producing cosmic rays come from outside the solar system, or at least are not generated by the sun. Results clearly show a uniform background of cosmic muons anti-correlated with solar activity. Whether data was collected on a cold winter day or a very hot summer afternoon, the count of muons remains fairly constant and around 500 per minute per unit area.

## Materials and Methods

### Detector and Detection Mechanism

The major components of the detector apparatus are the scintillator paddles and photomultiplier tubes. Both major components must be tested and calibrated for optimum performance. In a scintillator detector, a scintillating paddle is connected to a PMT via a light guide, and the entire apparatus is made light proof by way of photo-tape and photo-paper wrapping placed carefully over an aluminum foil layer. The absence of light leaks is tantamount to the reliability of scintillator-PMT detectors.

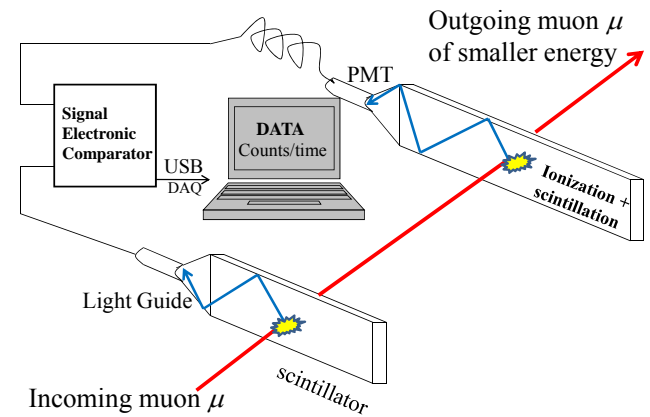

Figure 1: Components and Principle of detection. Mechanism: The detector is a set of two individual self-counting modules (PMT, scintillator and light guide) feeding data to an electronic comparator to isolate background noise from real muons, and to a USB data acquisition card.

The scintillator-PMT detectors will be used in pairs to detect muons. The setup uses two scintillators, mainly, to eliminate counting random events from electronic noise, in addition to giving the detector directionality. The detector operates under the assumption that an energetic muon travels near the speed of light, passing through two parallel scintillator paddles nearly simultaneously. This results in simultaneous generation of signals, which reach the electronic counter in nearly the same time. Those two signals are interpreted as a single coincident count. Using coincidence counting greatly reduces the chance that a large signal can be caused by an event other than the passage of an energetic muon.

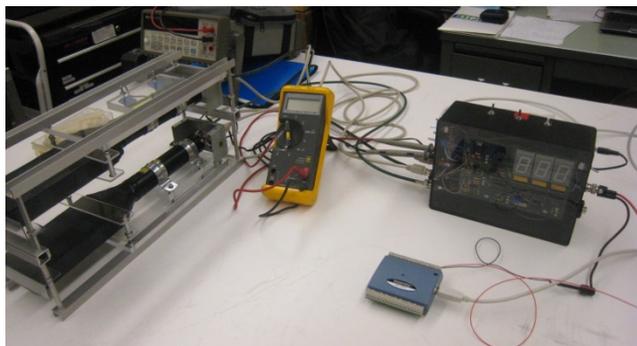

Figure 2: Cosmic muon detector. Two modules stacked on top of each other to detect muons travelling vertically downward. The count is displayed on a computer interface (not shown here) via a DAQ; the count is further duplicated by a standalone digital display of an electronic signal comparator.

Scintillators can be categorized into organic and inorganic types (Syed 2007). In organic materials, scintillation is due to excitation of electrons in a ground singlet state to a higher singlet or triplet level. This excitation is caused by an incident particle or wave imparting energy to the material via one of several modes of collision. An electron promoted to a higher energy level will decay to either the ground singlet state directly, emitting a photon quickly (fluorescence) or to an adjacent triplet level, which then decays to the ground triplet state. This process takes longer to complete and is called delayed fluorescence. When traversed by ionizing radiation, the scintillator's organic material emits electromagnetic radiation of such a wavelength that it absorbs the same emitted photons very efficiently. Thus, to ensure that scintillation light can be utilized for detection, organic scintillators are doped with a substance that will re-emit photons of a lower wavelength, often in the visible spectrum, so that the scintillating material is transparent to the new photons. Doping the scintillator material is a process that involves mixing a wavelength-shifting agent into the plastic solution, creating a homogenous mixture of different scintillating molecules, allowing for a uniform wavelength shifting.

A light guide connects the scintillator paddles with the PMTs and transmits light between them via total internal reflection (Collier and Wolfley 2006). PMTs are devices which convert incoming photons into electric signals with magnitude proportional to the incident photon flux. Upon striking the cathode of the PMT, a photon displaces an electron, via photoelectric effect; the electron is then accelerated through a series of dynodes through a strong electric potential. At each dynode, the electron is absorbed, and an increased number of electrons are emitted towards the next dynode due to the increasing electric potential at each dynode. This avalanche effect can create a signal at the anode with a magnitude of tens of millions of electrons. At the anode, the current signal is converted to a pulse of measurable voltage signal, which can be read on an oscilloscope. In our case, we connect the PMT-scintillator pair to an electronics module that can count the total number of pulses in a predefined time.

The electronics of the apparatus consist of a pulse counter and an optional DAQ module. The counter is responsible for providing power to the PMT and reading signals from the PMTs as counts. A gain voltage knob at the counter changes the power input of the PMT, thereby changing the sensitivity of the PMT. The plateauing of the PMTs, which indicates the optimal operating voltage, allows for greater sensitivity with little interference of background radiation. The collection of resistors, capacitors, diodes and integrated circuits compares PMT output pulses against a given voltage; this voltage difference can be as small as 7 mV can still make a detection event. Then the electronic circuit determines, through a series of logic gates, if the pulses from each PMT are close enough in time to call the pulses coincidences. To qualify as a count, the electronics require that hits in each module occur within a predetermined time of few milliseconds; only then will the electronic circuits advance the counter. The pulse counter can be connected to two detectors at the same time. A three-point toggle switch in the counter enables connection to either one of the counters separately or to both, in which case coincident counts are measured. The counter, however, is limited to making one-minute counts only. If counting for a longer period of time is desired, the DAQ module must be used. It connects the counter to a computer, and with necessary software, allows for counts ranging to any amount of time.

**Result**

*Flux Dependence on Elevation*

It has previously been determined that the flux of cosmic muons is proportional to altitude within a certain elevation range (Gaisser 1990). Thus, the feasibility of our radiation detection unit as a reliable portable muon-tagging device is dependent on its ability to confirm the accepted model for atmospheric muon flux. Because of the Lorentz length contraction, cosmic muons can travel farther than classically limited by their 2.2 μs lifetime. However, it stands to reason that the probability of a muon decaying increases the longer it travels. Thus, we expect a decrease in muon flux as distance increases downwards from the cosmic-shower atmospheric boundary.

To test the flux dependence on elevation, the standardized detector modules were placed at several locations at different elevations, keeping the parameters of the detector units constant for all data collections. The data was taken at each elevation to determine muon flux per unit area per unit time. Our findings were checked for correlation against accepted flux vs. elevation trends. The coincidence counts for a detection unit pair at each location is given in Fig. 3.

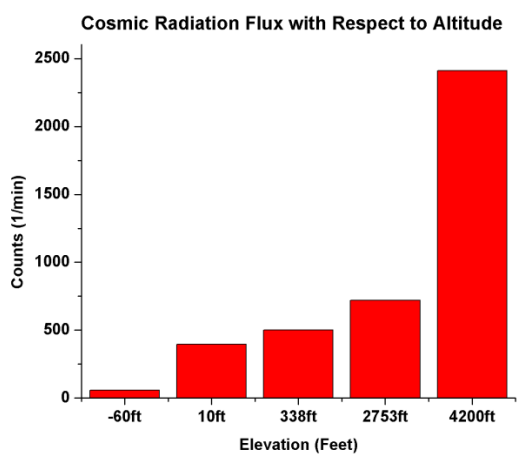

Figure 3: Flux as function of elevation. Locations tested included: Magnolia, AR (338ft), Galveston, TX (10ft), Mt. Magazine, AR (2753ft), Mt. Hamilton, CA (4200ft) and inside the Radiation Counting Laboratory (RCL) in the NASA Lyndon B. Johnson Space Center (- 60ft).

*Flux Dependence on Different Date of the Year*

The prevalent idea is that the flux of cosmic muons is unaffected by the temperature, solar activity and other climatic conditions on the earth. Cosmic muons are believed to be the result of the interaction of CR with the Earth's atmosphere. The precise origin of those CR is still unknown. Solar activity does not actually fall out of suspicion, and hence we have tried to determine if solar activity has any impact on the creation of cosmic muons. For this purpose, we took several one- minute coincidence-count measurements during different times of the year. The solar flux, sky conditions, and the temperature were noted. If the assumption that the solar activity is ineffective in the creation of cosmic muons is correct, there would not be any significant difference in the muon counts measured at different time of the year when the climatic conditions on Earth are different. In Fig. 4, we can clearly see that the counts measured per minute of the muons are not significantly different, and comparable. The small amount of difference can be attributed to the uncertainty in the measurements. Ten simultaneous measurements were taken, and an average of those measurements was taken on each date (Fig. 4, Table 1).

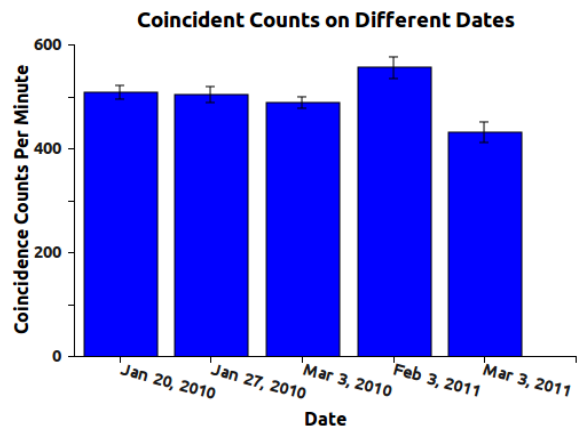

Figure 4: Flux as a function time and temperature. Irrespective of the day of the year or the temperature of the particular day, muon flux remains fairly uniform.

Table 1: Counts per minutes on different dates, over a one year span.

| Date | Counts | Uncertainty |
|---|---|---|
| Jan 20, 2010 | 507 | 23 |
| Jan 27, 2010 | 503 | 22 |
| Mar 3, 2010 | 487 | 22 |
| Feb 3, 2011 | 555 | 24 |
| Mar 3, 2011 | 458 | 21 |

**Discussion and Conclusion**

Muons are created in the upper atmosphere, approximately 50,000 feet above sea-level. Traveling at close to the speed of light, it takes them no more than 100 μs to reach sea-level. In the current paper, however, we measured a flux of about 1 muon per minute per square-cm at Galveston Bay and a flux of about 100 muons per minute per square-cm at Mt. Hamilton (4200 ft). Clearly not all muons reach sea-level; the larger the elevation, the more muons we registered. As charged particles, muons are subject to Coulomb interactions during their trip from the atmosphere into rocks on the Earth. In accordance with previous findings, this suggests that some of them are slow to start with, so they lose their kinetic energy and decay to an electron, an electron-antineutrino, and a muon-neutrino before reaching the Earth's surface. The highly energetic muons traverse matter and scatter through electromagnetic interactions; nevertheless, we did detect some of them 60 feet under the ground. Our investigation of the flux variation versus elevation shows a strong correlation between the two. The second important question that we addressed concerns the source of CR; can we identify a cosmological origin? Understanding the origin and acceleration mechanism of CR is a fundamental area of astroparticle physics with major unanswered questions. The uniform flux we observed over various times and days of the year clearly rules out the sun as a source. Furthermore, the number of counts did not vary with the time of day due to varying quantities of solar energy interfering with the travel of CR. Therefore, whatever their source, it must reside beyond our solar system or even our galaxy. This fact is not new; it has been long known, but with a fairly simple and portable detector we arrive at the same conclusions. The detector we describe in this manuscript has been a very useful tool in gaining an insight to the interesting world of muons.

## Acknowledgements

We would like to thank Arkansas Space Grant Consortium for providing funding. We are also grateful to Dr. Thomas Wilson at Johnson Space Center for fruitful discussion, and for giving us access to the radiation counting laboratory. Finally, many thanks are due to Dr. Azriel Goldschmidt at Lawrence Berkeley National Laboratory for his countless important suggestions and encouragement.